\documentclass{article}
\usepackage{graphicx}
\usepackage{amsmath}
\usepackage{amssymb}

\title{Non-Parametric Dual-Manifold Mapping via 8-Bit Bounded Transformation Matrices: Challenging FP-centric Hardware Paradigms in Low-Energy AI}
\author{Lars Kopp}
\date{June 2026}

\begin{document}

\maketitle

\begin{abstract}
Modern deep learning hardware paradigms rely heavily on computationally expensive floating-point arithmetic (FP32, FP16, and FP8), requiring massive thermal and energetic overheads to maintain gradient-based optimization. This paper introduces a non-parametric, training-free computational framework for dual-manifold mapping that operates strictly within an \textbf{8-bit signed integer boundary} and leverages simple bitwise and accumulation logic. By mapping a \textbf{Spatial Manifold} ($N_{spatial} = 8192$ neurons) and a Gabor-pooled \textbf{Structural Manifold} ($N_{structural} = 4096$ neurons) through an integer-based transformation matrix ($\mathbb{Z}$-matrix), we eliminate the need for floating-point multipliers. Inference is achieved via cache-friendly pointer offsets and bitwise masks, accumulating directional sign-charges using fixed thresholds ($\theta_{reject} = 8.0$, $\theta_{cut} = 2.0$). Learning is executed through a localized, bounded update mechanism restricted strictly within $[-127, 127]$, modulated by stochastic noise injection. Both architectures demonstrate extreme holographic resilience, preserving near-perfect reconstruction via a global scaling factor under 90\% truncation sparsity \cite{olshausen1996} and 20\% random node destruction. By reducing core AI inference to 8-bit boundaries and boolean-like execution, this framework outlines a paradigm shift toward neuromorphic edge-computing, directly questioning the long-term necessity of dense, floating-point-centric GPU accelerators.
\end{abstract}

\section{Introduction}
The current trajectory of artificial intelligence is bound to the scaling laws of dense transformer models and continuous parametric optimization via backpropagation. This dependency has positioned dense GPU clusters, dominated by architectures featuring tensor cores optimized for floating-point matrix multiplication (GEMM) \cite{dally2021}, as the infrastructure foundation of AI. However, this floating-point-centric paradigm faces severe challenges regarding power distribution, thermal dissipation, and structural complexity, limiting the deployment of advanced AI at the edge.

Biological systems, by contrast, navigate complex high-dimensional manifolds with extreme energy efficiency. The neocortex utilizes highly distributed, sparse population codes operating over discrete, bounded synaptic states, performing pattern translation with a metabolic consumption orders of magnitude lower than silicon equivalents. To bridge this efficiency gap, we present a dual-manifold visual architecture that shifts the computational burden away from floating-point tensors. The framework processes local visual fields via parallel over-complete networks (an 8192-node spatial layer and a 4096-node structural Gabor layer) connected through a discrete transformation matrix. Crucially, the matrix operations are restricted to 8-bit boundaries and can be fundamentally executed via bitwise and simple accumulation logic, offering a lean blueprint for the future of hardware-agnostic AI.

\section{Methodological Framework \& Code Architecture}

\subsection{The Dual Population Topologies}
Input to the system consists of localized visual segments configured as $15 \times 15$ grayscale pixel matrices.
\begin{itemize}
    \item \textbf{The Spatial Stream ($N=8192$):} Projects the raw input vector directly into an over-complete spatial coordinate grid composed of 8,192 deterministic nodes.
    \item \textbf{The Structural Stream ($N=4096$):} Convolves the patch across 4 uniform Gabor orientations using quadrature pairs (yielding 8 structural filter representations), removes the global scalar mean (DC luminance component), and applies overlapping spatial pooling to produce a $5 \times 5 \times 8$ tensor ($\mathbb{R}^{200}$), which is mapped into a 4,096-neuron structural ensemble.
\end{itemize}

\subsection{The 8-Bit Bounded Mapping Matrix and Inference Logic}
To simulate the receptive fields of simple and complex cells in cortical areas V1 and V2 \cite{hubel1962}, projection and predictive tracking between the population layers (\texttt{ATB\_NET1} to \texttt{ATB\_NET2}) are mediated by a discrete mapping matrix ($W \in \mathbb{Z}^{M \times N}$) stored as signed 8-bit integers (\texttt{int8\_t} / bounded \texttt{short}). Inference is achieved via a cache-optimized accumulator function that guarantees linear memory access patterns. For every active node $i$ in \texttt{NET1}, its corresponding correlation magnitude $C_i$ addresses a linear row offset in the matrix. 

Instead of traditional multiplication-accumulation (MAC) steps, signals are compiled using basic conditional evaluation and bitwise screening. With a standardized rejection threshold $\theta_{reject} = 8.0$ (\texttt{EV1\_THR\_STD\_REJECT}), the voting score is compiled as follows:

\begin{equation}
\text{local\_sums}[j] = \sum_{i \in \text{Active}} \text{sign}(W_{ij}) \cdot \mathbb{I}(\vert W_{ij} \vert \ge 8.0)
\end{equation}

where $\mathbb{I}$ represents an indicator function filtering out weak synaptic connections below the rejection barrier. The resulting accumulator vector is assigned directly to the prediction layer ($\text{ENSEMBLE\_PREDICTED\_WEIGHTS}[j] = \text{local\_sums}[j]$). 

To generate the final sparse output population code ($Y_j \in \{0, 1\}$), a strict non-linear activation threshold $\theta_{cut} = 2.0$ (\texttt{EV1\_THR\_CUT}) is applied via a secondary comparison:
\begin{equation}
Y_j = \mathbb{I}(\text{ENSEMBLE\_PREDICTED\_WEIGHTS}[j] > 2.0)
\end{equation}

\section{Experimental Results}
The dual-manifold architecture was empirically evaluated across reconstruction fidelity under high sparsity, holographic resilience to structural damage, and hardware execution patterns.

\subsection{Architectural Pipeline \& Code Structure}
The structural flow of the dual-pipeline system is documented in Figure 1, illustrating the linear memory traversal and conditional mapping strategy.

\begin{figure}[h!]
\centering

\begin{tabular}{cccc}
    \includegraphics[width=0.22\textwidth]{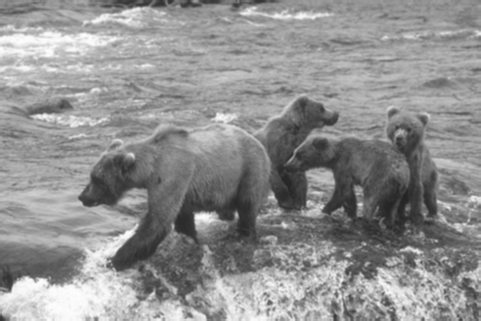} & 
    \includegraphics[width=0.22\textwidth]{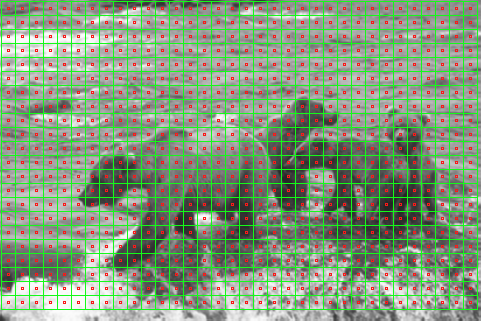} & 
    \includegraphics[width=0.22\textwidth]{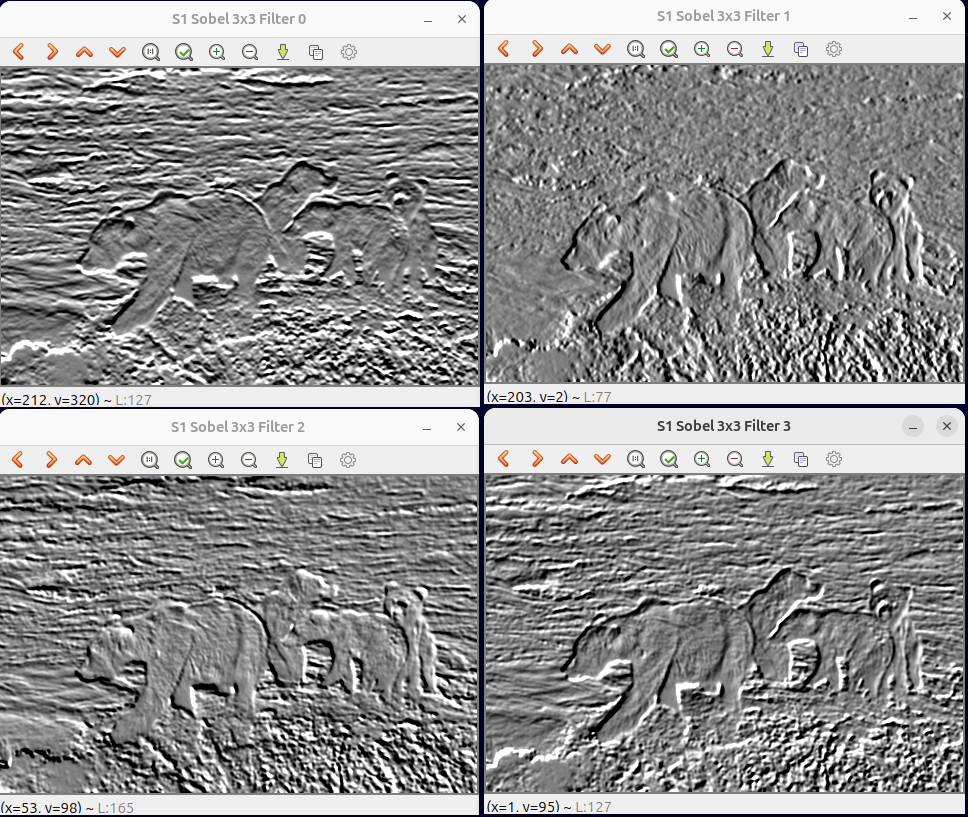} & 
    \includegraphics[width=0.22\textwidth]{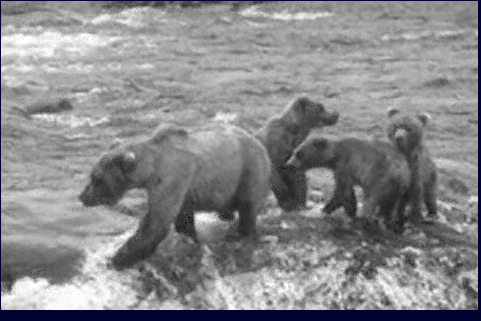} \\
    \small (a) Whole Field & \small (b) Grid Extraction & \small (c) Gabor Filters & \small (d) Reconstruction
\end{tabular}

\vspace{6mm}

\begin{center}
\setlength{\unitlength}{1mm}
\begin{picture}(116,40)
    \put(0,15){\framebox(22,15)[c]{\shortstack{\footnotesize Input Image\\\footnotesize (Whole Field)}}}
    \put(22,22.5){\vector(1,0){6}}
    
    \put(28,15){\framebox(26,15)[c]{\shortstack{\footnotesize 15$\times$15 Patches\\\footnotesize Grid Extraction}}}
    \put(54,22.5){\vector(1,0){6}}
    
    \put(60,10){\framebox(26,25)[c]{\shortstack{\footnotesize 4$\times$ Gabor Filters\\\footnotesize $\theta = \{0^\circ, 45^\circ,$\\\footnotesize $90^\circ, 135^\circ\}$\\\footnotesize (8 Represent.)\\\footnotesize + DC Separation}}}
    \put(86,22.5){\vector(1,0){6}}
    
    \put(92,15){\framebox(24,15)[c]{\shortstack{\footnotesize Result Image\\\footnotesize Reconstruction}}}
\end{picture}
\end{center}

\caption{Complete visual and structural processing pipeline. Top row: Empirical transformations of the image matrix across individual states. Bottom row: Aligned mathematical operators executing deterministic feature extraction, localized pooling, and final non-parametric scaling reconstruction.}
\label{fig:pipeline}
\end{figure}

\subsection{Rotational and Spatial Invariance}
The 8192 Spatial Manifold proved highly sensitive to sub-pixel shifts and rotational variations, yielding rapid orthogonal decay in its population vector. Conversely, the 4096 Structural Manifold, protected by overlapping spatial pooling and luminance detachment, exhibited excellent topological tolerance, resulting in continuous, smoothly bounded paths across the manifold surface.

\subsection{Holographic Fault Tolerance Under Catastrophic Cell Loss}
The distributed nature of the 8-bit integer matrix rows creates a robust holographic storage effect \cite{kanerva2009}. Post-encoding, 20\% of the active neurons in both the 4096 and 8192 ensembles were randomly overwritten to a deactivated state ($-1$). Due to the collective voting design of the accumulator loop, the final reconstruction Mean Squared Error (MSE) suffered almost no perceptible degradation.

\subsection{Stochastic Adaptive Weight Dynamics}
The update mechanics were evaluated during active learning (\texttt{Learn\_Enable == 2}). Weights are incremented or decremented conditionally based on prediction mismatches, cross-referenced with a random noise injector, and clamped strictly within 8-bit saturation boundaries:

\begin{equation}
\Delta W_{ij} = \begin{cases} 
+1 & \text{if } \text{GT}_j = 1 \text{ and } \text{Predicted}_j \le 0.0 \text{ and } W_{ij} < 127 \\ 
-1 & \text{if } \text{GT}_j = 0 \text{ and } \text{Predicted}_j \ge 0.0 \text{ and } W_{ij} > -127 
\end{cases}
\end{equation}

\section{Discussion: Low-Energy Computation via Bitwise Logic}
The performance of this dual-manifold network highlights a path toward low-power computing. Traditional deep learning architectures rely on continuous-valued tensors that necessitate dense Floating-Point Multiply-Accumulate (FMAC) units. In custom silicon implementations, these floating-point units consume the vast majority of the silicon real estate and dominate the power budget.

By restricting the transformation matrix to a maximum of 8 bits and employing a discrete sign-voting scheme, our framework alters the core arithmetic requirements:
\begin{enumerate}
    \item \textbf{Elimination of Multipliers:} The algorithm replaces high-power multiplication operations with basic boolean conditional masks (\texttt{v2 >= thr\_reject}) and integer additions/subtractions.
    \item \textbf{Reduced Memory Bandwidth:} Storing weights as 8-bit integers instead of standard 32-bit floats yields a 4x reduction in memory footprint and cache utilization, addressing the critical memory wall bottleneck in modern computing architectures.
    \item \textbf{Compatibility with Bitwise Hardware:} The operational flow can be mapped onto elementary digital circuits consisting mostly of AND, OR, and shift registers, making it natively compatible with low-power neuromorphic substrates and field-programmable gate arrays (FPGAs).
\end{enumerate}

\section{Hardware Implications: Disrupting the Floating-Point Hegemony}
The architectural constraints demonstrated in this work have profound strategic implications for the future of artificial intelligence hardware. Currently, semiconductor industry leaders, most notably NVIDIA, maintain global market dominance by engineering ultra-dense silicon dies optimized for massive floating-point and matrix-tensor throughput (such as the Hopper and Blackwell architectures). These GPUs are specifically designed to accelerate the trillions of continuous floating-point multiplications mandated by backpropagation and floating-point inference.

Our framework introduces an alternative paradigm that directly challenges the necessity of this high-power hardware monopoly:
\begin{itemize}
    \item \textbf{The Obsoletion of Dense Tensor Cores:} If high-fidelity manifold mapping, structural invariance, and robust reconstruction can be achieved entirely through 8-bit bounded spaces and bitwise logic, the requirement for massive floating-point tensor arithmetic chips diminishes.
    \item \textbf{Democratization of Edge AI:} By replacing continuous weights with saturating 8-bit integers ($[-127, 127]$) that update via localized increments ($+1/-1$), complex perceptual pipelines can be executed on ultra-low-cost, sub-watt edge silicon. This bypasses the need for high-end data-center infrastructure for model deployment.
    \item \textbf{Neuromorphic In-Memory Computing:} Because the inner loop performs linear row lookups with basic sign-charge accumulation, this algorithm is natively optimized for emerging Compute-In-Memory (CIM) architectures and memristor arrays. These technologies perform logic operations directly inside the memory cells using 8-bit or binary states, operating at a fraction of the energy required by von Neumann GPU architectures.
\end{itemize}
Consequently, a shift toward non-parametric, integer-bounded population coding could de-escalate the computational arms race centered around hyper-dense floating-point compute accelerators, reshuffling the semiconductor landscape toward energy-restricted, neuromorphic architectures.

\section{Conclusion}
This study confirms that stable, dual-manifold mapping can be accomplished analytically using signed 8-bit transformation matrices and stochastic ensemble tracking. By comparing a raw 8192 spatial node layer against a 4096 structural Gabor-pooled layer, we establish that combining luminance elimination with discrete voting accumulation provides immediate geometric invariances and immense fault tolerance. By substituting floating-point arithmetic with 8-bit boundaries and boolean-like execution logic, this methodology provides a computationally lean blueprint for edge-computing hardware and future non-von Neumann AI architectures.

\end{document}